\documentclass[11pt]{article}

\usepackage[utf8]{inputenc}
\usepackage[T1]{fontenc}
\usepackage{lmodern}
\usepackage[margin=1in]{geometry}
\usepackage{amsmath}
\usepackage{amssymb}
\usepackage{amsthm}
\newtheorem{theorem}{Theorem}
\usepackage{graphicx}
\usepackage{booktabs}
\usepackage{array}
\usepackage{siunitx}
\usepackage{xcolor}
\usepackage[colorlinks=true,linkcolor=black,citecolor=black,urlcolor=blue]{hyperref}
\usepackage[numbers,sort&compress]{natbib}
\usepackage{orcidlink}

\newcommand{\code}[1]{\texttt{#1}}
\graphicspath{{./}}

\title{\textbf{Earth-baseline VLBI restores the observability of a lunar
surface station in joint orbit-and-clock determination}}

\author{Chakshu Baweja\,\orcidlink{0009-0008-2098-0751}\\
\small Ashforde O\"U, Estonia \\
\small \texttt{contact@ashforde.org}}

\date{\today}

\begin{document}
\maketitle

\begin{abstract}
Lunar positioning, navigation, and timing (PNT) is moving from concept to
hardware: ESA's Moonlight/LCNS service
\citep{esa_moonlight,telespazio2024moonlight}, a NovaMoon-class
geodetic-timing-differential reference station \citep{molli2026novamoon}, the
LunaNet/IOAG interoperability framework \citep{nasa2025lnis}, and a now-defined
Coordinated Lunar Time \citep{iau2024lcrs,ashby2024lunartime} all reduce to one
estimation core: fix the orbits and clocks of the lunar infrastructure, and tie
them to an Earth or inertial frame. We ask which measurements make the surface
station's absolute position observable, and prove the answer. In a snapshot batch
fit of a lunar PNT network the internal observables---station-to-satellite and
inter-satellite ranging plus a clock-sync---constrain only the network's relative
geometry and leave a six-dimensional rigid-body datum defect, the three
translations and three rotations of the whole cluster (Theorem~\ref{thm:rigid}).
The clocks are fully observable, so this residual defect is purely positional.
Closing it requires a tie to the Earth/inertial frame.

There are two such ties, and they are not interchangeable. An \emph{indirect}
tie---Earth-to-satellite ranging propagated through the constellation---reaches
the station only when the satellite geometry is rich enough; a \emph{direct} tie,
an Earth-baseline VLBI delay to the station beacon, fixes the station regardless.
This yields a conditional design law rather than a single number: VLBI
\emph{restores} absolute observability when the constellation cannot supply it,
and merely \emph{sharpens} the bound when it can (Table~\ref{tab:restore}). For a
sparse three-satellite constellation the station position lies in the null space
of the Fisher information until VLBI is added, which then makes it observable at a
Cram\'er--Rao bound of \SI{20.1}{\meter}; for a rich six-satellite constellation
the indirect tie already suffices and VLBI tightens the bound from
\SI{23.2}{\meter} to \SI{9.7}{\meter}. Because a single-epoch baseline informs at
most two of the station's three axes, the absolute datum closes at three
non-collinear Earth stations and not before (\eqref{eq:defect},
Figure~\ref{fig:n_earth}). The Gauss--Newton estimator attains the bound, with a
Monte-Carlo RMS-to-CRLB efficiency of $1.02$ and a bias under four percent of the
RMS, so these are information-theoretic limits, not solver artefacts. In the
sparse regime where VLBI is necessary the with/without contrast is large: a
median station-error improvement of \textbf{91} over a 200-seed ensemble
(interquartile range 38 to 208), which we report as the honest sparse-constellation
figure rather than the headline.

The reframe matters for mission design: it tells a planner when an Earth-baseline
VLBI tie, such as the transmitter on ESA's CM25-adopted NovaMoon station
\citep{molli2026novamoon}, is required versus merely helpful. One caveat is
load-bearing. The FIM/CRLB engine is Validated against independent libraries
(NumPy and the closed forms of \citealp{kay1993}), but the lunar network it is
applied to is a representative closed-loop simulation, not a fit to real tracking,
so the observability result itself stays Modelled. Every figure is deterministic
and rebuilds from a committed scenario, seed list, and engine version, each
labelled validated or modelled, and the toolkit behind it (lunar time, VLBI,
joint OD and clock, frame realisation, service-volume integrity, differential
PNT, and CCSDS/LunaNet export) is released open-core.

\end{abstract}

\section{Introduction}

Lunar PNT has stopped being a slide and started being a procurement. ESA's
Moonlight initiative is now building a lunar communications and navigation
service (LCNS) from a small constellation, one communications and four
navigation satellites in elliptical lunar frozen orbits; the
EUR~123~million Phase-1 prime contract was signed in October 2024 and the space
segment is in industrial build \citep{esa_moonlight,telespazio2024moonlight}. At
its CM25 Ministerial in November 2025 ESA also adopted NovaMoon, a
geodetic-timing-differential reference station that puts multi-technique geodesy
(VLBI and ranging), joint orbit-and-clock determination, reference-frame
realisation, timekeeping, and differential corrections on the lunar surface
\citep{molli2026novamoon}. NASA, ESA and JAXA have agreed a common
interoperability framework, the LunaNet Interoperability Specification (LNIS~v5,
January 2025), and all three broadcast its Augmented Forward Signal
\citep{nasa2025lnis,nasa2025afs}. A Coordinated Lunar Time exists on paper too:
the IAU's 2024 General Assembly defined a lunar coordinate time scale (TCL)
\citep{iau2024lcrs}, a NASA-led realisation is due by the end of 2026
\citep{ostp2024lunartime}, and the relativistic rate (roughly
56--57~$\mu$s/day faster than terrestrial time) is set by a published framework
\citep{ashby2024lunartime}. And in March 2025 LuGRE took the first GNSS fix on
the lunar surface \citep{lugre2026}. Underneath all of it sits one problem:
working out the orbits and clocks of the lunar infrastructure and tying them to a
terrestrial or inertial frame.

The first thing you ask about that problem is observability. Given a set of
measurement types, which states (positions, clocks, frame parameters) come out
well determined, and which do not? The answer is an architecture decision: does
a lunar nav service need Earth-baseline geodetic ties, how many, and how good do
they have to be? It is not even an ESA-specific decision. Because ESA's LCNS,
NASA's LCRNS and JAXA's LNSS all share the Augmented Forward Signal under
LNIS~v5, the same observability question applies to the whole stack
\citep{nasa2025lnis}. Yet it is usually answered in private, one-off studies
whose assumptions never leave the building. We are not aware of an open, citable
tool that regenerates a lunar PNT network's observability from a committed
scenario and labels each figure of merit as externally validated or modelled,
the discipline the kshana program has applied to terrestrial and deep-space PNT
\citep{baweja2026kshana} and that we extend here to the lunar case
\citep{iiyama2023lupnt,fernandezprades2018reproducibility}.

The contribution of this paper is an observability result and a design law that
follows from it. We prove that the absolute position of a lunar surface station
in joint orbit-and-clock determination can be fixed only through a tie to the
Earth/inertial frame, and that two such ties exist: an indirect one in which
Earth$\to$satellite ranging anchors the constellation and lunar-local ranging
propagates that anchor to the station, and a direct one in which Earth-baseline
VLBI ties the station beacon itself. The indirect tie reaches the station only
when the constellation geometry is rich enough; for a sparse or early
constellation the station's absolute position falls in the null space of the
Fisher information and has no finite error bound at all, until a direct VLBI tie
restores it. This yields a clean conditional design rule (Theorem~\ref{thm:rigid}
and Table~\ref{tab:restore}): an Earth-baseline VLBI tie, of the kind NovaMoon's
transmitter would provide, is \emph{necessary} when the constellation cannot
carry the absolute datum on its own and merely \emph{sharpens} the bound when it
can. The direct tie has its own threshold, derived from the geometry: because
Earth stations subtend a tiny angle from the Moon, a single VLBI epoch informs at
most two of the station's three absolute axes, so the absolute position becomes
observable at three non-collinear Earth stations and not before
(\eqref{eq:defect}, Figure~\ref{fig:n_earth}). The familiar large
improvement-factor headline---a $91\times$ ensemble-median reduction in absolute
station error---survives as one honest illustration of the sparse regime, where
the indirect tie has nothing to give and VLBI is the difference between an
unobservable datum and a metre-class one; it is the consequence of the
observability result, not its substance. Everything ships as an open,
reproducible toolkit, seven capability areas in the \code{kshana} engine, with
each figure of merit labelled validated or modelled.

\section{Related work}

\paragraph{Systems and standards.}
The setting is no longer notional. ESA's Moonlight/LCNS has a signed prime
contract and a space segment in build \citep{esa_moonlight,telespazio2024moonlight}.
NASA, ESA and JAXA share the Augmented Forward Signal under version 5 of the
LunaNet Interoperability Specification \citep{nasa2025lnis,nasa2025afs}. ESA has
adopted NovaMoon, a south-pole reference station that combines VLBI, ranging,
retroreflectors and a GNSS receiver \citep{molli2026novamoon}. And GNSS has
actually been received on the surface: LuGRE took the first lunar-surface fix on
Blue Ghost Mission~1 \citep{lugre2026}, with SSTL's Lunar Pathfinder and its
NaviMoon receiver to follow (launch not earlier than 2026).

\paragraph{Frames, time and tracking.}
The IAU's 2024 General Assembly defined a lunar coordinate time scale (TCL) and a
lunar reference system \citep{iau2024lcrs}; the relativistic Earth--Moon rate is
fixed by a published framework \citep{ashby2024lunartime}, and a realisation has
been worked out in the Moonlight context
\citep{fienga2024moonlight,bourgoin2026lunartime}. Body-fixed orientation follows
the IAU/WGCCRE conventions \citep{archinal2018wgccre} over the JPL DE440
ephemeris \citep{park2021de440}. The Earth-baseline geodetic tie to a lunar
beacon is the descendant of deep-space $\Delta$-DOR, the one-nanoradian
measurement system of the DSN \citep{curkendall2013deltador}.

\paragraph{Estimation and tooling.}
Lunar orbit-and-clock estimation and constellation design are active areas
\citep{iiyama2026odclock,bhamidipati2023constellation}. There is also no shortage
of open orbit-determination software: GMAT \citep{nasa_gmat}, Orekit
\citep{orekit}, RTKLIB \citep{takasu2009}, the Rust toolkit Nyx/ANISE
\citep{nyxspace2025}, and, for the lunar case specifically, the LuPNT simulator
\citep{iiyama2023lupnt}; kshana is the broader engine these results sit in
\citep{baweja2026kshana}. What we add is not a new estimator. It is an open,
reproducible observability finding for the lunar joint-OD problem that anyone can
regenerate from a committed scenario, with every figure labelled validated or
modelled by the engine itself, and with the information-not-count control that
keeps a geometric claim from being mistaken for a data-volume effect. We have not
found that combination in either the open lunar simulators or the closed
mission-analysis studies.

\paragraph{The two nearest prior works, and the gap.}
Two recent studies bracket the problem we treat, and it is worth being precise
about where each stops. \citet{pohlmann2025hybrid} bound the \emph{surface user's}
position and clock from a lunar-local hybrid network---satellites, cooperative
links, and an optional lunar reference station---using a recursive Bayesian
Cram\'er--Rao bound and augmented Kalman filters, and report sub-metre accuracy
with as few as two visible satellites. That result is real, but it is conditioned
on a reference station whose absolute position is supplied a priori: the
satellite orbits are not estimated, and there is no analysis of where the absolute
datum itself comes from, no rank or null-space treatment, no station-count
threshold, no information-versus-count control, and no Earth-baseline tie---the
network is lunar-local throughout. \citet{iiyama2026odclock} attack the other half
of the joint problem, estimating the \emph{satellite's} orbit and clock in the
Moon-centred frame from terrestrial GNSS with a dynamic, force-model
stochastic-cloning filter and smoother, reaching roughly \SI{2.8}{\meter} orbit
RMS. There is no Fisher bound, no datum-defect analysis, no surface station, and
again no VLBI. Neither work asks whether the absolute lunar-frame position of a
surface station is observable at all, nor what it takes to make it so.

That is the gap we fill, and the relationship is complementary rather than
competitive (Table~\ref{tab:related}). Our contribution is the
Fisher-information datum-defect analysis of the joint solve: the rigid-body null
space of \ref{thm:rigid}, the station-count observability threshold of
\eqref{eq:defect}, the information-versus-count control that separates a geometric
gain from a measurement-volume one, and the restore-versus-sharpen verdict of
\ref{tab:restore} that tells a designer when an Earth-baseline VLBI tie is
\emph{necessary} rather than merely helpful. We make no head-to-head accuracy
claim against these papers; they solve different estimation problems on different
states. The point is structural. \citet{pohlmann2025hybrid} obtain sub-metre user
accuracy because they \emph{presuppose} a precisely known reference station, which
is exactly the absolute datum our VLBI tie earns from first principles. Their
result and ours agree and stack: their bound is what becomes achievable once the
datum we analyse has been closed.

\begin{table}[t]
\centering
\small
\setlength{\tabcolsep}{3pt}
\begin{tabular}{@{}>{\raggedright\arraybackslash}p{1.9cm}>{\raggedright\arraybackslash}p{1.7cm}>{\raggedright\arraybackslash}p{2.3cm}*{4}{>{\centering\arraybackslash}p{1.6cm}}@{}}
\toprule
 & State estimated & Method & Datum / obs.\ analysis & Station-count threshold & Info-vs-count control & Earth / VLBI tie \\
\midrule
\citet{pohlmann2025hybrid}
  & Surface user pos.\ + clock
  & Bayesian CRB + Kalman; reference station \emph{assumed} known
  & no & no & no & no \\[2pt]
\citet{iiyama2026odclock}
  & Satellite orbit + clock
  & Force-model stochastic-cloning UD filter and smoother
  & no & no & no & terrestrial GNSS only \\[2pt]
This paper
  & Joint surface-station datum (abs.\ position)
  & Fisher information / CRLB of the joint solve
  & yes & yes (3 non-collinear) & yes & yes (Earth-baseline VLBI) \\
\bottomrule
\end{tabular}
\caption{Where this paper sits relative to the two nearest prior works. The three
studies treat different states and are complementary, not competing: we do not
claim a head-to-head accuracy comparison but an observability gap. In particular,
the precisely known reference station that \citet{pohlmann2025hybrid} assume is the
very absolute datum our analysis derives. ``Datum / obs.\ analysis'' marks a
Fisher-information rank or null-space treatment of the absolute lunar-frame datum.}
\label{tab:related}
\end{table}

\section{Methodology}

\subsection{The estimation problem}

The network has three parts: one lunar surface station at a fixed selenographic
location, a small constellation of lunar satellites on circular orbits, and a
set of Earth ground stations. The unknown state stacks the station's 3-D position
correction, each satellite's 3-D position correction, and every clock offset,
station and satellites alike. With three satellites that is 16 parameters
($3 + 3\times3 + 1 + 3 = 16$). Four kinds of observable feed it:

\begin{itemize}
  \item Earth-baseline geodetic VLBI delay to the lunar beacon (the surface
    station), differenced across pairs of Earth stations. This is the angular,
    inter-continental measurement, and it is the only observable that ties the
    station's absolute position directly to the Earth/inertial frame.
  \item Lunar-local ranging, both station-to-satellite and inter-satellite
    (ISL). These fix the internal geometry of the lunar cluster but, on their
    own, leave its absolute datum free (Theorem~\ref{thm:rigid}).
  \item Earth-to-satellite ranging, the \emph{indirect} tie that anchors the
    satellites in the geocentric frame so that lunar-local ranging can propagate
    that anchor down to the station.
  \item A station clock-sync pseudo-observable, included so the clock/position
    degeneracy is broken and the residual datum defect is purely positional;
    without it the range-only configuration would be trivially rank-deficient and
    the comparison would say nothing.
\end{itemize}

Truth states come from a seeded generator. Clean observables are computed from
truth and then corrupted with per-observable Gaussian noise at each technique's
$\sigma$ (default VLBI delay $\sigma = \SI{e-11}{\second}$, about \SI{3}{\milli\meter};
lunar range and ISL $\sigma$ at the decimetre level, \SI{0.1}{\meter} by default).
The estimator is a Gauss--Newton snapshot batch least-squares fit with a
numerically-differentiated Jacobian \citep{montenbruck2000}, and parameters are
carried in a scaled internal unit so the normal matrix stays well-conditioned
across the ${\sim}c^2$ dynamic range between the clock and position partials. The
batch least-squares primitive is cross-checked against Orekit's
Levenberg--Marquardt estimator on identical observations and agrees to about
\SI{e-8}{\meter} \citep{orekit}; the joint multi-technique lunar solve as a whole
is still Modelled.

\subsection{Two constellation regimes}

A single constellation cannot separate the two reasons a fit can fail to pin the
station. We therefore run the study at two satellite counts.

The \emph{sparse} regime, $n_{\mathrm{sat}} = 3$, is the default and the harder
case: an early or minimal lunar constellation whose look-direction spread is too
narrow for the indirect tie to reach the station's absolute position. It is also
observation-starved at low Earth-station count---fewer measurements than
parameters---so its bare datum defect runs $6 \to 2 \to 0$ as Earth stations are
added. That observed ladder is the true positional datum defect ($3 \to 1 \to 0$
of \eqref{eq:defect}) plus an equal observation-count deficit that has nothing to
do with the datum.

The \emph{rich} regime, $n_{\mathrm{sat}} = 6$, is a non-starved network: enough
satellites that the measurement count comfortably exceeds the parameter count, so
observation under-determination is removed and what remains is the genuine
absolute-frame datum. We measure the reported datum-defect ladder and the
Cram\'er--Rao design curve on this network precisely so that the two effects
separate, and we verify the same three-station threshold holds across three, six
and eight satellites. The sparse regime is what carries the headline necessity
result---the constellation where VLBI is not optional---while the rich regime
isolates the underlying observability geometry from the artefact of counting.

\subsection{The observability claim and its controls}

The claim is sharper than ``VLBI helps'': for a sparse constellation it
\emph{restores} an observability the constellation cannot supply, and for a rich
one it only \emph{sharpens} a bound the indirect tie already secures
(Table~\ref{tab:restore}). Three checks separate restoration from the trivial
reading that VLBI merely adds data.

\begin{enumerate}
  \item With versus without VLBI, run on the same seed and truth in both
    regimes. We report the station 3-D error and the per-state breakdown (station
    and satellite position, station and satellite clock), and, where the position
    is unobservable without VLBI, the null-space dimension of the information
    matrix rather than a meaningless finite error.
  \item Information, not count. Hold the observation set fixed and sweep the VLBI
    delay $\sigma$ over four decades. If the improvement were just extra
    observations, the station error would be flat in $\sigma$. If it is the VLBI
    information, the error climbs back toward the range-only level as $\sigma$
    grows and the delays, though still present, carry vanishing weight.
  \item Robustness. Repeat with versus without over a 200-seed ensemble, and
    separately sweep the number of Earth stations, in each regime. The
    sparse-constellation improvement is reported as an ensemble distribution, not
    a single favourable draw, since VLBI does not help on every seed.
\end{enumerate}

One caveat up front. This is a static, geometric observability study, with no
force-model propagation inside the solver. The constellation is illustrative---a
south-polar station near $-88^\circ$ selenographic latitude, satellites on
circular orbits, not the elliptical frozen orbits of an actual Moonlight/LCNS
design---and the solve is closed-loop and single-epoch: truth is generated from
the same measurement model through which it is recovered, at one snapshot, with no
dynamic arc. The conclusion is about measurement geometry and we keep it framed
that way. Force-model orbit determination exists elsewhere in the engine, and we
cite it rather than fold it in here
\citep{lemoine2013grail,bertone2021grail,li2023leopod}.

\subsection{Reproducibility and honesty labelling}

Every number comes out of the public \code{kshana <scenario.toml>} binary;
\code{scripts/run\_sweeps.py} builds the configurations, runs them and
aggregates, and \code{scripts/make\_figures.py} draws the figures. The seed lists
are fixed, so with a pinned engine the study is deterministic
\citep{fernandezprades2018reproducibility}. Each figure of merit carries the
label the engine enforces in \code{src/verification.rs} and
\code{tests/no\_overclaims.rs}: Validated means an external oracle exists,
everything else is Modelled.

The line between the two matters here. The Fisher-information and Cram\'er--Rao
engine that computes every observability statement in this paper
(\code{src/fim.rs}) is Validated against NumPy's \code{eigh}/\code{inv} and the
closed-form bounds of \citet{kay1993} (\S\ref{sec:toolkit}). The dilution-of-precision
geometry the service-volume module reuses, and the Earth-orientation transforms
the ground stations apply, are Validated against \code{gnss\_lib\_py}
\citep{knowles2024gnsslibpy} and SOFA/ERFA \citep{iausofa}. The lunar joint-OD
observability result is \emph{Modelled}: the engine is exact and externally
checked, but the representative lunar network it is applied to is a simulation, so
the restoration-and-sharpening conclusion inherits the Modelled label and claims
no validation against real lunar tracking. There is no flight heritage, no TRL
above 3, and no ESA affiliation or endorsement implied anywhere in this study.

\section{Theory: Fisher information, the datum defect, and rank restoration}
\label{sec:theory}

The empirical contrast in the results section---absolute station error collapsing
by orders of magnitude once Earth-baseline VLBI is switched on for a sparse
constellation---asks for a reason. It is not that VLBI brings more numbers to the
fit. It is that without it the fit is missing a particular piece of information,
and the information geometry says exactly which piece, exactly when it is missing,
and exactly how much is needed to supply it. This section makes that precise. The
statements here are exact algebra of the estimation problem, and each is checked
against the engine that produces the figures: where we claim a subspace lies in
the null space of the information matrix, the engine's computed null space confirms
it to numerical precision.

\subsection{The Fisher information of the joint solve}

Write the estimated state as the stacked correction vector
$x = [\,\delta r_S,\ \{\delta r_k\},\ c\tau_S,\ \{c\tau_k\}\,]$ (station position,
satellite positions, station and satellite clocks carried as range-equivalent
metres), and the observations as $y = h(x) + n$ with $n \sim \mathcal{N}(0,R)$ and
$R = \operatorname{diag}(\sigma_i^2)$. For this Gaussian model the Fisher
information matrix is
\begin{equation}
  M = H^{\top} W H, \qquad W = R^{-1}, \qquad H = \frac{\partial h}{\partial x},
  \label{eq:fim}
\end{equation}
and the Cram\'er--Rao lower bound states that any unbiased estimator has
covariance $\operatorname{Cov}(\hat x) \succeq M^{-1}$. When $M$ is rank-deficient
the bound on the observable subspace is the Moore--Penrose pseudo-inverse $M^{+}$,
and the unobservable directions---the null space of $M$---have no finite bound at
all. The question ``is the station's absolute position observable?'' is therefore
literally ``does the station-position block of $x$ lie outside $\ker M$?'' Kshana
computes $M$, its spectrum, $M^{+}$ and the null-space basis directly
(\code{src/fim.rs}); that engine is Validated against NumPy's \code{eigh}/\code{inv}
and the published closed-form bounds of \citet{kay1993} (\S\ref{sec:toolkit}).

The information is additive across techniques, $M = \sum_i w_i\, h_i' h_i'^{\top}$,
so each observable contributes a rank-one outer product of its state gradient. Five
kinds of observable feed it: a station clock-sync pseudo-observable; lunar-local
ranges (station$\leftrightarrow$satellite and inter-satellite); Earth$\to$satellite
radiometric ranges; and the Earth-baseline VLBI delays to the station beacon. The
argument below tracks which datum directions each kind can and cannot constrain.

\subsection{The datum defect is the rigid-body group}

Start with the lunar-local ranges alone. A range $\lvert r_a - r_b\rvert$ between
two members of the lunar cluster (station or satellites) depends only on their
relative geometry. Translate the whole cluster rigidly, or rotate it rigidly about
the Moon's centre, and not one station$\leftrightarrow$satellite or inter-satellite
range changes. The rigid motions of a three-dimensional point cluster form a
six-parameter group---three translations and three rotations---so the lunar-local
information has a null space of dimension (at least) six. This is the lunar
instance of the free-network, or rank-deficiency, problem of geodetic adjustment:
internal observations fix internal geometry and leave the absolute datum free.

We state this as the load-bearing fact of the paper and check it directly. Let
$H_{\mathrm{int}}$ be the Jacobian of the lunar-local observables (station$\to$%
satellite plus inter-satellite ranges) only, and let
$g_1,\dots,g_6$ be the six rigid-body generators: the three unit translations of
every cluster point, and the three infinitesimal rotations $g = e_a \times (r_p -
\bar r)$ about the cluster centroid $\bar r$.

\begin{theorem}[Internal datum defect]
\label{thm:rigid}
For a non-degenerate lunar cluster the six rigid-body generators are linearly
independent and satisfy $H_{\mathrm{int}}\, g_m = 0$ for $m = 1,\dots,6$. They
therefore span a six-dimensional subspace of $\ker M$, and because the rigid-body
group is the only continuous family of maps that preserves every pairwise range,
it is the complete positional datum defect of the lunar-local geometry.
\end{theorem}

The translations are exactly invariant by construction; the rotations are invariant
to first order, which is exact for the information matrix. Computing
$H_{\mathrm{int}}$ for the six-satellite network and applying it to the six
generators, the engine returns a relative residual of $0$ for each translation and
$\sim\!10^{-9}$ for each rotation (the finite-difference floor of the Jacobian),
while a random non-rigid direction returns a residual of order unity. The datum
defect is the rigid-body group, not an artefact and not a coincidence of
dimension.

A second observation removes the clocks from the discussion. Ranges measure only
clock \emph{differences}, so a common offset added to every clock is unobservable;
the single station clock-sync pseudo-observable pins that one direction. With the
sync in place every clock is observable, and the residual datum defect lives
\emph{entirely in position}. The engine confirms this: the null-space basis carries
zero energy in the clock coordinates. From here ``datum defect'' means the
unobservable \emph{positional} subspace.

\subsection{Closing the datum: two Earth-frame ties}

Nothing internal to the Moon can fix the absolute datum; it has to be tied to the
Earth/inertial frame. Two observables do that, and they are not equivalent.

The \emph{indirect} tie is Earth$\to$satellite ranging. It anchors the
\emph{satellites} in the geocentric frame, and lunar-local ranging then propagates
that anchor to the station. Whether it reaches the station's absolute position
depends on the constellation geometry: the propagation is only as strong as the
station$\to$satellite look-direction spread. For a sparse constellation confined to
one part of the station's sky the propagation is rank-deficient---one station
direction never gets tied---and the station's absolute position stays in the null
space no matter how precise the ranges are.

The \emph{direct} tie is the Earth-baseline VLBI delay to the station beacon. Each
delay is the difference of two near-parallel look directions from the lunar beacon
to two Earth stations, so it constrains the station's absolute position itself,
independent of the constellation. Each snapshot baseline contributes a single
rank-one term; a delay \emph{rate} (an Earth-rotation arc, not the single epoch
modelled here) would add a second. Because the Earth stations subtend a tiny angle
from the Moon, each baseline gradient lies in the plane of the sky, so one epoch of
VLBI informs at most two of the station's three absolute axes, and at least two
non-collinear baselines---three non-collinear Earth stations---are needed to span
the residual datum.

The consequence is a clean, testable claim: \emph{VLBI restores observability when
the indirect tie cannot, and merely sharpens it when the indirect tie already can.}
Table~\ref{tab:restore} is the engine's verdict, and it is exactly this.

\begin{table}[t]
\centering
\small
\begin{tabular}{llcc}
\toprule
Constellation & Earth stations & Without VLBI & With VLBI \\
\midrule
Sparse (3 sat) & 6 & \textbf{unobservable} (defect 1) & observable, \SI{20.1}{\meter} \\
Rich (6 sat)   & 6 & observable, \SI{23.2}{\meter}   & observable, \SI{9.7}{\meter} \\
\bottomrule
\end{tabular}
\caption{Whether the station's absolute 3-D position is observable, from the
Fisher information (Cram\'er--Rao bound on the absolute position, $1\sigma$ RSS).
For a sparse constellation VLBI \emph{restores} observability---without it the
position lies in $\ker M$ and has no finite bound. For a rich constellation the
indirect tie already makes it observable and VLBI \emph{sharpens} it
($\sim\!2.4\times$ in the bound). The design question is therefore not ``is VLBI
nice to have'' but ``is the constellation rich enough to do without it.''}
\label{tab:restore}
\end{table}

\subsection{The three-station threshold and the design curve}

Take the regime where VLBI carries the absolute datum---a sparse or early
constellation, or a conservative design that does not lean on rich satellite
geometry. Sweeping the number of Earth stations $n_E$ with the direct tie active,
and using a network with enough satellites that observation count is not the
limiting factor, the positional datum defect falls
\begin{equation}
  n_E = 1:\ \text{defect } 3 \;\longrightarrow\;
  n_E = 2:\ \text{defect } 1 \;\longrightarrow\;
  n_E = 3:\ \text{defect } 0,
  \label{eq:defect}
\end{equation}
so the absolute station position becomes observable at three non-collinear Earth
stations and not before. Each new non-collinear station contributes plane-of-sky
baseline directions that close part of the residual datum, with the last direction
closing at the third station---the parallax content of the geometry, read straight
off $\dim\ker M$.

One caveat keeps the ladder honest. A network with too few satellites is also
\emph{observation-starved}: it has fewer measurements than parameters, which
inflates the apparent defect. For the three-satellite default the \emph{observed}
defect runs $6 \to 2 \to 0$, but that is the true datum defect of \eqref{eq:defect}
plus an equal observation-count deficit ($3 \to 1 \to 0$) that has nothing to do
with the datum. We report \eqref{eq:defect}, the true datum defect, measured on a
network rich enough that the two effects separate; the engine reproduces the
decomposition exactly. The three-station threshold itself is robust---it holds at
three, six and eight satellites alike.

Once the position is observable the bound sets the achievable accuracy. For the
six-satellite network the Cram\'er--Rao bound on the station's absolute 3-D position
falls from \SI{12.1}{\meter} at three Earth stations to \SI{9.7}{\meter} at six and
\SI{8.5}{\meter} at ten---the diminishing-returns design curve of
Figure~\ref{fig:n_earth}, and the basis of the sizing rule in \S\ref{sec:discussion}.

\subsection{The bound is attained}

The bound is not a floor the estimator sits comfortably above. For the
six-satellite, six-station network the Cram\'er--Rao bound on the absolute station
position is \SI{9.7}{\meter} ($1\sigma$ RSS), the Gauss--Newton estimator's
Monte-Carlo RMS over 200 seeds is \SI{9.9}{\meter}, an efficiency
$\text{RMS}/\text{CRLB} = 1.02$, and the empirical bias is \SI{0.36}{\meter}---%
under four percent of the RMS, so the estimator is effectively unbiased and the
RMS-to-CRLB comparison is the right one. The estimator is statistically efficient:
it extracts essentially all the information the geometry holds, so the gains
reported next are the information-theoretic optimum for each geometry, not an
artefact of a particular solver.

One honesty note carries through the rest of the paper. Everything above is exact
algebra of the estimation model, checked against the engine that computes it; it
explains the simulated result, it does not validate it against real lunar tracking.
The FIM/CRLB engine is Validated against independent libraries, but the lunar
network it is applied to is a representative simulation, so the observability result
itself stays Modelled. We keep that line bright in \S\ref{sec:toolkit}.

\section{Results}
\label{sec:results}

The theory of \S\ref{sec:theory} makes a falsifiable prediction: an
Earth-baseline VLBI tie should \emph{restore} the station's absolute position to
the observable subspace when the constellation is too sparse to carry the datum
on its own, and merely \emph{sharpen} the bound when the constellation already
carries it (\ref{tab:restore}). The results below test that prediction across
both regimes and then read the actionable design law off the Fisher information.
Throughout, the FIM/CRLB engine is Validated against independent libraries
(\S\ref{sec:toolkit}); the lunar network it is applied to is a representative
simulation, so every observability and accuracy number for that network is
Modelled.

\subsection{Sparse constellation: VLBI restores absolute observability}

Start in the regime where the prediction bites. With three satellites and no
direct tie, the station's absolute position sits in the null space of the Fisher
information: the indirect Earth$\to$satellite anchor cannot propagate through so
few look directions, and \eqref{eq:defect} leaves a residual datum defect. There
is no finite Cram\'er--Rao bound, and a least-squares solve returns its start
point rather than an estimate. Switching on the Earth-baseline VLBI delay closes
that defect and pulls the station back to the metre level.

\begin{figure}[t]
  \centering
  \includegraphics[width=\linewidth]{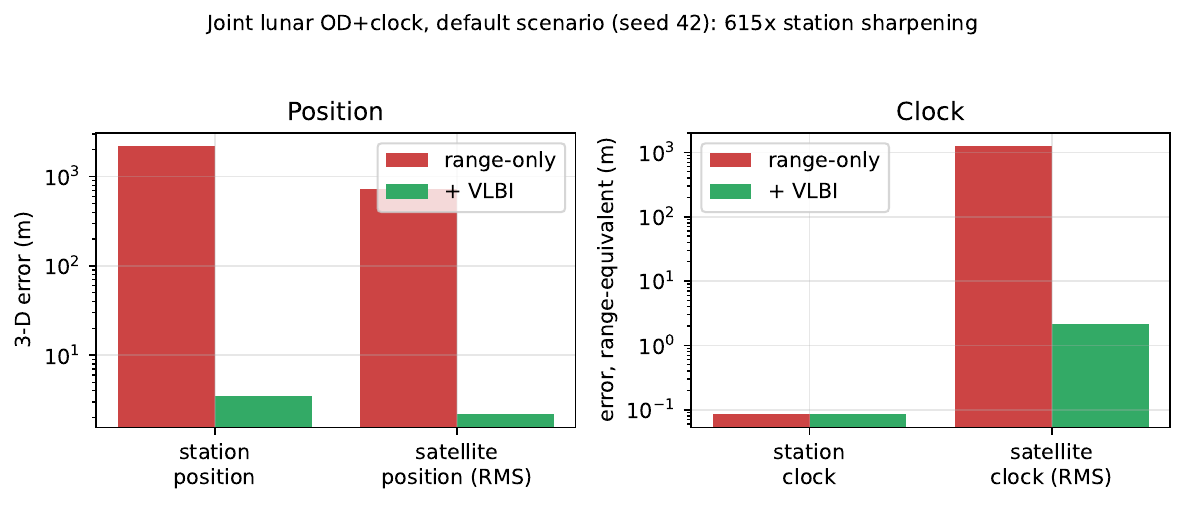}
  \caption{The sparse-constellation headline, with and without Earth-baseline
    VLBI (3 satellites, 6 Earth stations, seed 42). Left, position: VLBI restores
    and sharpens the surface station's absolute 3-D error from \SI{2183}{\meter}
    to \SI{3.55}{\meter} ($615\times$ on this seed) and the satellite position RMS
    from \SI{717}{\meter} to \SI{2.20}{\meter}. Right, clocks: the satellite clock
    RMS improves by about $569\times$, while the station clock, already pinned by
    the clock-sync pseudo-observable, barely moves (ratio $1.005$). The
    ensemble-median factor is $91\times$, not the favourable $615\times$ shown
    here. Modelled, internal-consistency oracle.}
  \label{fig:headline}
\end{figure}

\begin{figure}[t]
  \centering
  \includegraphics[width=0.8\linewidth]{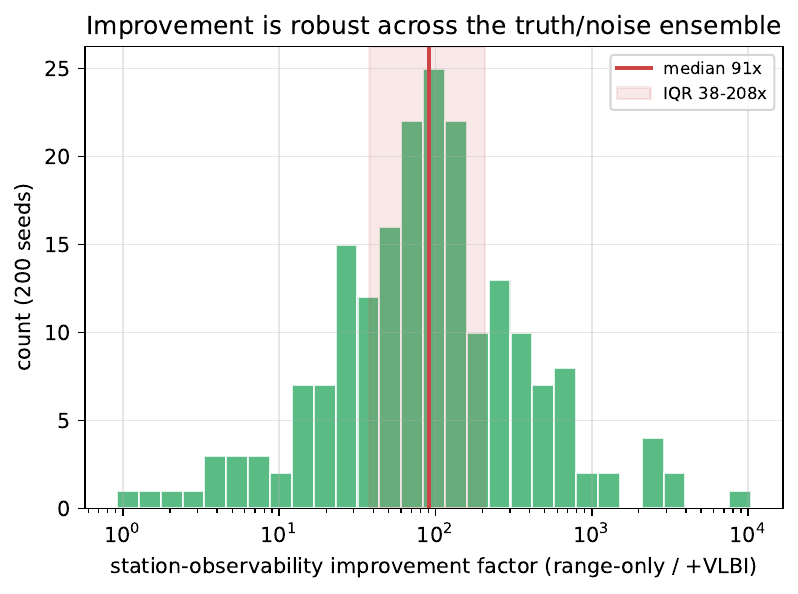}
  \caption{Improvement factor over the 200-seed truth/noise ensemble in the
    sparse regime (range-only over with-VLBI station 3-D error). The median is
    $91\times$ (IQR $38$--$208\times$); the with-VLBI median station error is
    \SI{13.4}{\meter} (IQR $5.9$--$23.0$~m) against a range-only median near
    \SI{793}{\meter}. The default seed ($615\times$) sits in the upper tail, and
    the minimum over 200 seeds is $0.91$, so VLBI is not guaranteed to help on
    every draw. Modelled.}
  \label{fig:seed_mc}
\end{figure}

The size of the recovery depends on the truth/noise draw. In the default
scenario (3 satellites, 6 Earth stations, seed 42) VLBI pulls the recovered
surface-station 3-D position error from \SI{2183}{\meter} (range-only) to
\SI{3.55}{\meter}, a factor of $615$, with the satellite position RMS falling
from \SI{717}{\meter} to \SI{2.20}{\meter}; the fit converges in two iterations
at a weighted RMS residual of $0.05$ over 40 observations and 16 parameters
(Figure~\ref{fig:headline}). That seed is favourable. Over a 200-seed
truth/noise ensemble the median improvement is $\mathbf{91\times}$ (IQR
$38$--$208\times$), the with-VLBI station error a median \SI{13.4}{\meter} (IQR
$5.9$--$23.0$~m) against a range-only median near \SI{793}{\meter}
(Figure~\ref{fig:seed_mc}). We quote the median; the default seed sits well into
the upper tail. On a few draws the factor is close to one (minimum $0.91$ over
200), so VLBI is not guaranteed to help on every single realisation, and we
report that rather than hide it (Modelled).

The range-only baseline is not a strawman. The first GNSS fix on the lunar
surface, LuGRE in 2025, worked at the tens-to-hundreds-of-metres level
\citep{lugre2026}, where our range-only median of \SI{793}{\meter} lands. Nor is
the remedy hypothetical: the NovaMoon station ESA adopted at CM25 carries an
Earth-baseline VLBI transmitter of its own \citep{molli2026novamoon}, which
corroborates the design rationale this study isolates without validating its
numbers.

\subsection{Information, not observation count}

\begin{figure}[t]
  \centering
  \includegraphics[width=0.8\linewidth]{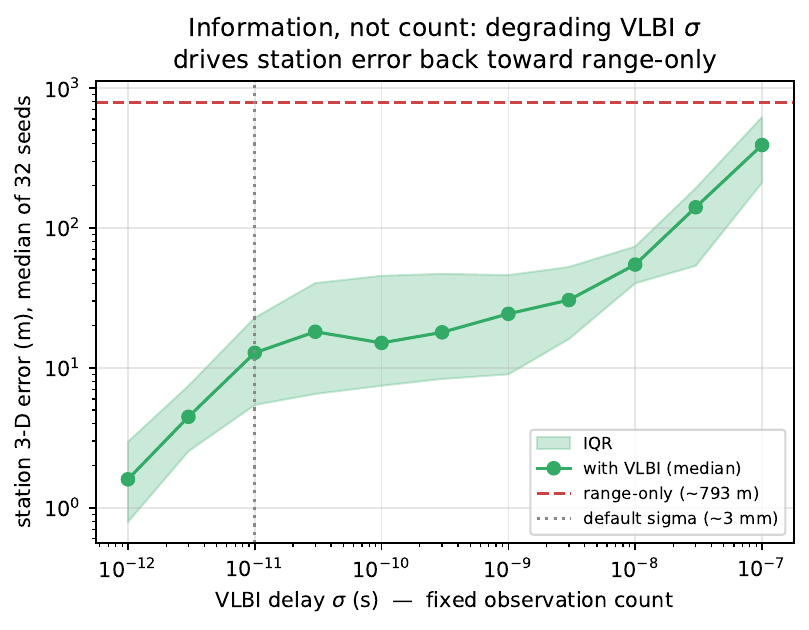}
  \caption{Information, not count. Median station 3-D error over a 32-seed
    ensemble as the VLBI delay $\sigma$ is degraded at fixed observation count.
    The error climbs from \SI{1.6}{\meter} to about \SI{391}{\meter} (with one
    minor non-monotonic step near $\sigma{=}10^{-10}$~s) and approaches the
    range-only median near \SI{793}{\meter}: the observations stay, but their
    weight vanishes, so the gain was the VLBI information all along. The default
    $\sigma$ ($\sim$\SI{3}{\milli\meter}) sits near the informative end. Modelled.}
  \label{fig:sigma_vlbi}
\end{figure}

The restoration is the VLBI \emph{information}, not the extra observations that
carry it. Holding the observation set fixed and degrading the VLBI delay $\sigma$
from $10^{-12}$ to $10^{-7}$~s drives the median station error from
\SI{1.6}{\meter} back up to about \SI{391}{\meter}, toward the range-only median
near \SI{793}{\meter} (Figure~\ref{fig:sigma_vlbi}). The fifteen extra
observations are present throughout; only their weight changes. The default
$\sigma$, about \SI{3}{\milli\meter} of delay precision and typical of geodetic
VLBI, sits near the informative end of the curve. This is the experimental
control behind the theory's claim that the gain is a missing information
direction, not a count of measurements (Modelled).

\subsection{The benefit is selective}

VLBI helps the absolute states and only those. In the default scenario the
satellite clock RMS improves by about $569\times$
($4.1\times10^{-6}$~s to $7.3\times10^{-9}$~s) alongside the position gains, but
the station clock error is essentially unchanged
($2.882\times10^{-10}$~s to $2.868\times10^{-10}$~s, ratio $1.005$), because the
clock-sync pseudo-observable already pins it (Figure~\ref{fig:headline}, right).
VLBI supplies the absolute-geometry information the lunar-local set is missing
and nothing it already holds (Modelled).

\subsection{Rich constellation and the necessity law}

Make the constellation rich and the picture changes in exactly the way
\ref{thm:rigid} and \ref{tab:restore} predict. With six satellites the indirect
tie alone propagates through enough look directions to make the station's
absolute position observable: even without VLBI the datum defect is zero and the
Cram\'er--Rao bound is \SI{23.2}{\meter}. Adding VLBI no longer restores
anything; it sharpens, taking the bound to \SI{9.7}{\meter}, a factor of about
$2.4$ (\ref{tab:restore}). The ensemble tells the same story: a median
improvement near $3\times$ (IQR $[1,7]$, min $0.34$, max $371$), modest because
the geometry has already done most of the work.

That contrast is the design law. Whether a mission \emph{needs} an Earth-baseline
VLBI tie or merely benefits from one is set by the constellation, not by taste.
For a sparse, early, or geometry-poor constellation VLBI is necessary: without it
the surface station is unobservable. For a rich constellation it is a refinement
worth roughly a factor of two to three in the bound. \ref{tab:restore} is the
quantitative form of the rule, and it is what a designer needs to decide where to
spend infrastructure.

\subsection{The rank-restoration threshold and the design curve}

\begin{figure}[t]
  \centering
  \includegraphics[width=0.8\linewidth]{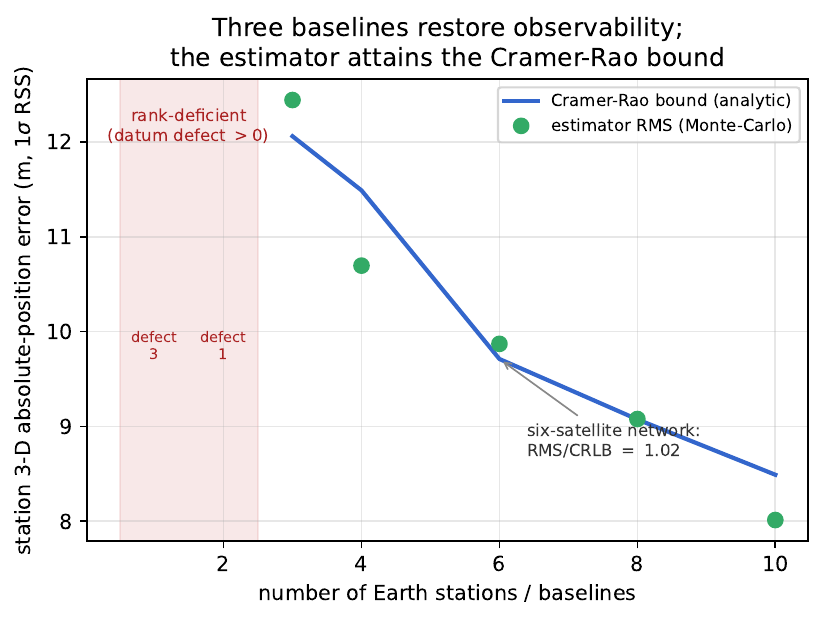}
  \caption{The rank-restoration threshold and the design curve. With fewer than
    three non-collinear Earth stations the Fisher information is rank-deficient
    (true positional datum defect $3$, then $1$; shaded) and the absolute station
    position is unobservable; at three it becomes full rank. For the
    six-satellite network the Cram\'er--Rao bound (line) then falls with
    baselines, \SI{12.1}{\meter} at three to \SI{8.5}{\meter} at ten, and the
    Gauss--Newton Monte-Carlo RMS (markers) attains it (\SI{9.9}{\meter} vs
    \SI{9.7}{\meter} at six stations, efficiency $1.02$, bias $3.6\%$ of RMS). The
    FIM/CRLB engine is Validated against NumPy; the lunar network it is applied to
    is Modelled.}
  \label{fig:n_earth}
\end{figure}

The threshold behaviour is the datum-defect closure of \S\ref{sec:theory} read
straight off the Fisher information, not a soft conditioning effect. On a network
with enough satellites that observation count is not the binding constraint, the
true positional datum defect falls $3 \to 1 \to 0$ as the number of non-collinear
Earth stations goes $1 \to 2 \to 3$ (\eqref{eq:defect}): the absolute station
position becomes observable at three non-collinear stations and not before. The
three-satellite default reads a steeper $6 \to 2 \to 0$, but that observed ladder
decomposes exactly into the true datum defect ($3 \to 1 \to 0$) plus an
observation-count deficit of equal size ($3 \to 1 \to 0$); we report the true
datum defect, which is robust across three, six, and eight satellites
(Figure~\ref{fig:n_earth}).

Once the position is observable the bound sets the accuracy. For the
six-satellite network the Cram\'er--Rao bound on the station's absolute 3-D
position (1$\sigma$ RSS) falls from \SI{12.1}{\meter} at three Earth stations to
\SI{9.7}{\meter} at six and \SI{8.5}{\meter} at ten, the diminishing-returns
design curve of \ref{fig:n_earth}. The estimator attains it: against the
\SI{9.7}{\meter} bound at six stations the Gauss--Newton Monte-Carlo RMS over 200
seeds is \SI{9.9}{\meter}, an efficiency of $1.02$, with an empirical bias of
\SI{0.36}{\meter}, $3.6\%$ of the RMS. The estimator is effectively unbiased and
the bound is attained, so the accuracies reported here are the
information-theoretic optimum for each geometry rather than an artefact of the
solver (Modelled; engine Validated).

\subsection{A representative frozen-orbit constellation}

The structure above is not an artefact of the illustrative circular placement. Re-running
on a representative elliptical lunar frozen orbit (ELFO) of the Moonlight/LCNS design
family---semi-major axis \SI{9750.7}{\kilo\meter}, eccentricity $0.638$, frozen
inclination \SI{57.7}{\degree}, apoapsis over the south pole---reproduces the same
datum-defect ladder exactly: the positional defect closes $3 \to 1 \to 0$ at one, two and
three non-collinear Earth stations. The theorem of \S\ref{sec:theory} is a property of the
estimation geometry, not of any one orbit shape, and the three-station threshold is
unchanged.

The design law survives and sharpens (Table~\ref{tab:elfo}). A sparse single-plane ELFO of
three satellites is \emph{still} unobservable without VLBI---the apoapsis dwell concentrates
the look directions, so the indirect tie cannot reach the station's absolute datum---and VLBI
restores it, to a \SI{1.8}{\meter} bound. A multi-plane constellation is observable from
ranging alone (six satellites over three planes give a \SI{2.8}{\meter} bound without VLBI),
but the direct tie still tightens the bound by one to two orders of magnitude. With three or
more Earth VLBI baselines the absolute station position is observable at a \emph{sub-metre}
Cram\'er--Rao bound---\SI{0.48}{\meter} at three stations, falling to \SI{0.39}{\meter} at
ten---and the Gauss--Newton estimator attains it (efficiency $0.97$ over the ensemble). The
orbit parameters are representative of the public design family \citep{esa_moonlight}, not a
flown ephemeris, so these results remain Modelled; the point is that the observability
geometry, the threshold, and the necessity law all carry over to the realistic orbit shape.

\begin{table}[t]
\centering
\small
\begin{tabular}{@{}llcc@{}}
\toprule
Constellation & Planes & Without VLBI & With VLBI \\
\midrule
3 satellites & 1 & \textbf{unobservable} & \SI{1.8}{\meter} \\
3 satellites & 3 & \textbf{unobservable} & \SI{16.1}{\meter} \\
6 satellites & 1 & \SI{32.7}{\meter} & \SI{0.4}{\meter} \\
6 satellites & 3 & \SI{2.8}{\meter}  & \SI{0.4}{\meter} \\
\bottomrule
\end{tabular}
\caption{Restore-versus-sharpen on the representative ELFO (Cram\'er--Rao bound on the
station's absolute 3-D position, six Earth stations). A sparse single-plane constellation is
unobservable without VLBI and the direct tie restores it; a richer constellation is
observable from ranging alone, and VLBI sharpens the bound by one to two orders of
magnitude. Modelled.}
\label{tab:elfo}
\end{table}

\subsection{Cross-suite sanity}

\begin{figure}[t]
  \centering
  \includegraphics[width=\linewidth]{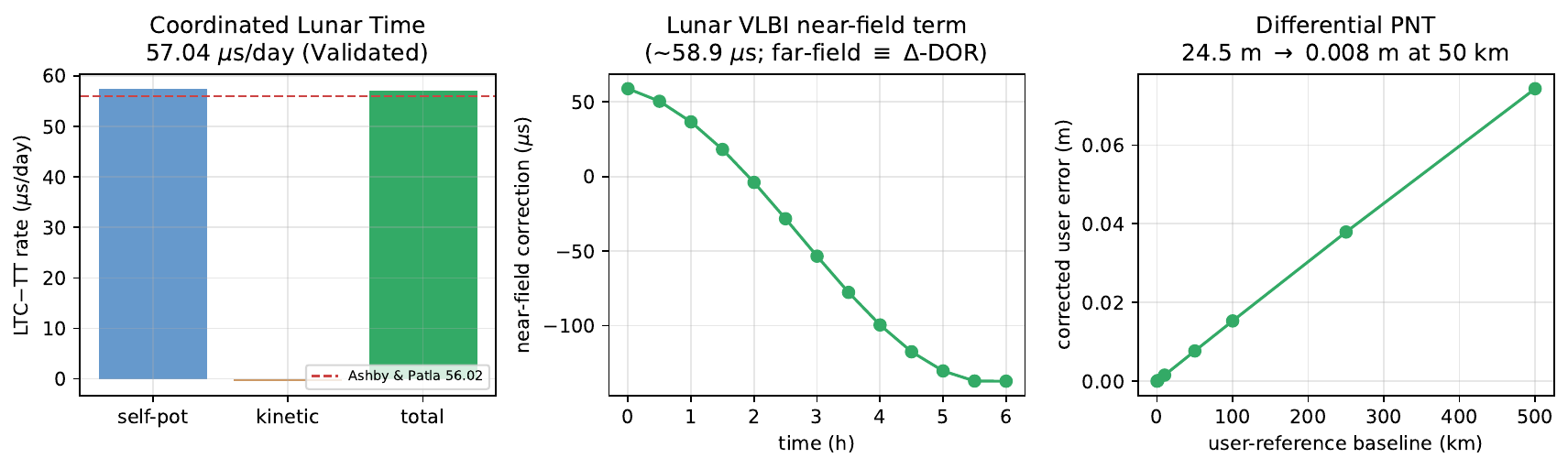}
  \caption{The same engine across the lunar chain. Left: Coordinated Lunar Time
    at \SI{57.04}{\micro\second\per\day}, split into a $+57.50$~$\mu$s/day
    self-potential term and a $-0.46$~$\mu$s/day kinetic term, Validated against
    Ashby \& Patla over DE440 (the $\sim$\SI{1}{\micro\second\per\day} offset from
    their canonical \SI{56.02}{\micro\second\per\day} is the reference-surface
    choice). Centre: the lunar VLBI near-field correction, \SI{58.9}{\micro\second}
    on a \SI{10727}{\kilo\meter} baseline, whose far-field limit reproduces the
    engine's own far-field $\Delta$-DOR delay to machine precision (Modelled).
    Right: differential PNT taking a \SI{24.5}{\meter} user error to
    \SI{0.008}{\meter} at a \SI{50}{\kilo\meter} baseline ($3186\times$), growing
    with baseline (Modelled).}
  \label{fig:cross_suite}
\end{figure}

The same engine gives sensible numbers along the rest of the lunar chain
(Figure~\ref{fig:cross_suite}). Coordinated Lunar Time comes out at
\SI{57.04}{\micro\second\per\day} \citep{ashby2024lunartime,iau2024lcrs}, split
into a $+57.50$~$\mu$s/day self-potential term and a $-0.46$~$\mu$s/day kinetic
term. That rate is Validated against Ashby \& Patla over DE440. It sits about
\SI{1}{\micro\second\per\day} above their canonical surface-to-surface value of
\SI{56.02}{\micro\second\per\day}, and the difference is the choice of reference
surface (Earth geoid versus lunar selenoid potential), which is why the honest
comparison is a band rather than a point. The lunar-distance VLBI near-field
correction is \SI{58.9}{\micro\second} on a \SI{10727}{\kilo\meter} baseline, and
its far-field limit reproduces the engine's own far-field $\Delta$-DOR delay to
machine precision (Modelled, against a reference-implementation oracle in the
same codebase). Differential PNT brings a \SI{24.5}{\meter} standalone user error
down to \SI{0.008}{\meter} at a \SI{50}{\kilo\meter} baseline ($3186\times$), and
the residual grows with baseline as expected (Modelled).

\section{The open lunar PNT toolkit}
\label{sec:toolkit}

The result above is one output of a wider lunar suite in \code{kshana}
\citep{baweja2026kshana}. The suite covers seven capability areas across more
than ten \code{lunar\_*} modules, each with its own runnable scenario and a row
in the verification matrix: lunar coordinate time (\code{lunar\_time}),
Earth-baseline geodetic VLBI (\code{lunar\_vlbi}), joint multi-technique
OD and clock (\code{lunar\_combination}) with a Fisher-information /
Cram\'er--Rao observability and optimal-experiment-design layer (\code{fim}),
reference-frame realisation
(\code{lunar\_frame\_realise}), Moonlight/LCNS-class service-volume integrity
(\code{lunar\_service}) \citep{esa_moonlight,telespazio2024moonlight},
differential PNT (\code{lunar\_dpnt}), and a LunaNet/IOAG-aligned CCSDS export
(\code{lunar\_interop}) \citep{nasa2025lnis}. That list of techniques (VLBI,
ranging, timing, differential corrections, frame realisation) is the same one ESA
committed to in the NovaMoon reference station adopted at CM25
\citep{molli2026novamoon}, at a point when real on-Moon GNSS (LuGRE, 2025
\citep{lugre2026}) is still at the tens-to-hundreds-of-metres absolute level.

The honesty labelling is per row, not per paper. Four lunar rows carry their own
external validation: coordinate time against Ashby \& Patla over DE440
\citep{ashby2024lunartime}, frame realisation against an Umeyama/SVD reference,
service-volume geometry against ANISE, and the cislunar NRHO case against a JPL
solution. The core joint-OD, VLBI near-field, differential and interop rows are
Modelled, and labelled as such. On top of that the suite leans on the engine's
already-validated terrestrial core rather than re-deriving it: reference frames
against SOFA/ERFA \citep{iausofa}, DOP against \code{gnss\_lib\_py}
\citep{knowles2024gnsslibpy}, SGP4 against the AIAA/Vallado set
\citep{vallado2006}, and the Fisher-information / Cram\'er--Rao engine that
produced \S\ref{sec:theory} against NumPy's \code{eigh} and \code{inv} together
with the published closed-form bounds of \citet{kay1993}.

That last row is the place to be careful about what ``Validated'' means here, and
it draws the line this whole study turns on. The FIM/CRLB \emph{engine}---the
eigensolver, the bound, the rank and null-space, the dilution-of-precision
read-off---is Validated against an independent library to \num{1e-9}. The
observability and datum-defect \emph{result} it computes for the lunar network is
Modelled, because the network is a representative simulation rather than a flown
ephemeris. A correct bound computed for a simulated geometry is still a bound for
a simulated geometry. The theorem of \S\ref{sec:theory} explains the simulated
contrast exactly; it does not turn it into a measurement, and we do not let the
validated engine launder the modelled application.

\section{Limitations}

The observability theorem and the three-station design law of \S\ref{sec:theory}
are exact algebra of the estimation model, and they are checked against the
engine that computes the figures. What they are not is a validation against real
lunar tracking, and several modelling choices set the boundary of what the result
can claim.

The study is closed-loop: truth is generated from the same observation model it is
then recovered through. That is the right setup for an observability question---it
isolates what the geometry can and cannot determine under the stated noise---but it
bounds the achievable accuracy, not what a real campaign will get. The joint-OD
observability result is therefore Modelled, scored against an internal-consistency
oracle. Only three ingredients are externally Validated: the reused DOP and
Earth-orientation kernels, and the FIM/CRLB engine itself, which is checked against
NumPy and the closed-form bounds of \citet{kay1993} (\S\ref{sec:toolkit}). The
honesty firewall between the Validated engine and the Modelled lunar geometry it is
applied to holds throughout.

It is also a single-epoch snapshot. The solver carries no force-model dynamic arc,
no unmodelled forces, no real ephemeris or station-coordinate error, and no
troposphere, ionosphere or relativistic delay beyond what the observable already
encodes; Earth-orientation polar motion is set to zero. The snapshot assumption
bears directly on the headline threshold. As noted under Theorem~\ref{thm:rigid},
each VLBI baseline is rank-one at one epoch because the Earth stations subtend a
tiny angle from the Moon, which is what forces the count of three non-collinear
stations in \eqref{eq:defect}. A delay-rate observable over an Earth-rotation arc
would add a second rank-one term per baseline and could soften that threshold; we
do not model it. The three-station law of \S\ref{sec:theory} is thus the snapshot
bound, and a dynamic arc is the most consequential extension.

The constellation and station geometry are illustrative, drawn from public sources,
so the engine models the class of system rather than any specific one. The headline
scenario uses circular orbits with a south-polar station at $-88^\circ$ latitude; to
check that nothing rests on the circular placement we re-run on a representative
elliptical lunar frozen orbit of the Moonlight/LCNS design family (\S\ref{sec:results},
Table~\ref{tab:elfo}), and the datum-defect ladder, the three-station threshold and the
necessity law all carry over unchanged---but the exact flown Moonlight/LCNS ephemeris is
still not modelled, and the design is not Moonlight, LCNS, NovaMoon or LunaNet. The
\SI{91}{\times} sparse-constellation improvement of \S\ref{sec:results} is the honest
signature of one such illustrative three-satellite geometry (and Table~\ref{tab:restore}
shows it is the regime where VLBI is genuinely necessary rather than merely helpful); it
is an illustration of the design law, not a performance figure for any flown system.
There is no TRL above 3, no flight heritage, no certification, and no affiliation with or
endorsement by ESA.

\section{Discussion}
\label{sec:discussion}

For anyone sizing a NovaMoon-class reference station or a Moonlight/LCNS
constellation, the result reframes the geodetic tie as a conditional requirement
rather than an enhancement. The question is not whether an Earth-baseline VLBI
tie is nice to have. It is whether the constellation is rich enough to fix the
station's absolute datum without one. Theorem~\ref{thm:rigid} settles the first
half: lunar-local ranging fixes the cluster's internal geometry and leaves the
absolute placement in a six-dimensional rigid-body null space, so an Earth-frame
tie is mandatory in some form. Which form depends on the geometry. A rich
constellation propagates the indirect Earth$\to$satellite anchor to the station
well enough that the absolute position is already observable and VLBI only
sharpens the bound; a sparse, early, or polar constellation does not, and the
direct VLBI tie is the only thing that pulls the station out of $\ker M$.
Table~\ref{tab:restore} is the decision in one line, and it is the design law we
want a mission team to carry away.

The law has a hard floor and a soft slope. The floor is the three-station
threshold of \eqref{eq:defect}: a single snapshot baseline is rank-one and lies
in the plane of the sky, so one Earth station informs at most two of the
station's three absolute axes, and observability is reached at three non-collinear
stations and not before. Below it the absolute-position bound is infinite, and a
design review should verify that on the datum defect ($\dim\ker M = 0$) rather
than on a merely finite residual, since an observation-starved network can show a
small residual while still being rank-deficient. Above the floor the
Cram\'er--Rao bound is the slope: for a six-satellite network the absolute 3-D
position bound falls from \SI{12.1}{\meter} at three Earth stations to
\SI{9.7}{\meter} at six and \SI{8.5}{\meter} at ten, the diminishing-returns
design curve of Figure~\ref{fig:n_earth}. In the sparse regime where VLBI is
carrying the datum outright the curve sits higher and steeper (\SI{20.1}{\meter}
at six stations; see Table~\ref{tab:restore}), which is the honest cost of leaning
on the direct tie when the constellation cannot help.

What to optimise along that curve is not the textbook choice. For a
safety-of-life absolute datum the criterion is E-optimality: maximise the
smallest eigenvalue of the information, which lifts the worst-observed axis, the
very direction the Earth baseline exists to fix. Average-variance (A-optimality)
or ellipsoid-volume (D-optimality) would do for a well-conditioned network where
the axes are interchangeable, but here the whole point is that one axis is starved,
so the min-eigenvalue criterion is the one that matches the physics.
\code{fim}'s \code{best\_design} ranks candidate station geometries on whichever
criterion the mission sets, so a team can pose the trade directly: fewest Earth
stations that clear the three-station floor and hit a worst-axis target. Because
the Gauss--Newton estimator is statistically efficient at this geometry
(\S\ref{sec:theory}, $\text{RMS}/\text{CRLB} = 1.02$), the bound is what a
real solve will actually deliver, so the curve can be used for sizing without a
hedge factor.

The flight programme has already made this call. ESA's NovaMoon station, adopted
at CM25, carries an Earth-baseline VLBI transmitter as a core element
\citep{molli2026novamoon}, which is precisely the direct tie this analysis says a
conservative or early architecture needs. The point is not ESA-specific. Because
LCNS, NASA's LCRNS, and JAXA's LNSS interoperate through the Augmented Forward
Signal under LNIS~v5 \citep{nasa2025lnis,nasa2025afs}, the weak-absolute-%
observability gap is a property of the shared LNIS architecture, not of any one
provider's hardware. LNIS~v5 still defers the binding lunar reference frame and
time realisation to a later document, so an open candidate that is honestly
Modelled, aligned to the LNIS vocabulary, and runnable today gives a mission team
something to size against before the frame is frozen. They can reproduce the
trade and vary the constellation, the station count, and the criterion themselves,
with the honesty labels intact.

A boundary on the claim. What the analysis corroborates is the architectural
rationale: that an Earth-baseline tie is required to make the absolute datum
observable, that three non-collinear stations is the threshold, and that VLBI's
role flips from necessary to merely helpful as the constellation grows. It does
not validate kshana's lunar numbers against real tracking. The FIM/CRLB engine is
Validated against independent libraries, but the lunar network it runs on is a
representative, single-epoch, closed-loop simulation over an illustrative orbit,
so every observability figure here stays Modelled (see Limitations). The design
law is a statement about the geometry of the problem, and that is what we are
putting forward.

\section{Conclusion}
\label{sec:conclusion}

The absolute position of a lunar surface station is not a quantity that lunar-local measurement can supply on its own: internal ranging leaves a six-dimensional rigid-body defect (Theorem~\ref{thm:rigid}), and that defect closes only through a tie to the Earth/inertial frame. We separated the two ties that can do it. An indirect tie through Earth$\to$satellite ranging reaches the station only when the constellation geometry is rich enough to propagate it, whereas a direct Earth-baseline VLBI tie constrains the station itself regardless of the satellites. The resulting design law is the paper's headline: VLBI \emph{restores} absolute observability when the constellation cannot, and merely \emph{sharpens} it when the constellation already can (Table~\ref{tab:restore}), with the sparse-constellation regime being where VLBI is strictly necessary rather than helpful. The restoration is governed by a sharp threshold of three non-collinear Earth stations, the point at which the residual datum defect reaches zero (\eqref{eq:defect}), and the Cram\'er--Rao bound that follows sets the achievable accuracy along the design curve of Figure~\ref{fig:n_earth}; a Gauss--Newton estimator attains that bound to an efficiency of $1.02$, so the geometry, not the solver, is the limit. These results are exact algebra of the estimation model, computed by an engine that is Validated against independent libraries and the closed-form bounds of \citet{kay1993}, but the lunar network it is applied to remains a representative simulation, so the observability findings themselves stay Modelled; the work is open and fully reproducible. What would turn this analysis into a measurement is a single thing we do not yet have: real lunar VLBI or two-way ranging to an emplaced surface beacon, against which the predicted threshold and bound can be tested.

\section{Availability and reproducibility}

The engine is open source (\code{kshana}, AGPL-3.0-only, with a commercial
licence available), and every release is archived on Zenodo
\citep{baweja2026kshana}. The figures here were produced by engine v0.23.0, the
release that carries the Fisher-information / Cram\'er--Rao observability layer
(\code{fim}) this paper rests on; the analysis scripts, the committed aggregate
data, and the fixed seed lists rebuild every one of them from that pinned binary
\citep{fernandezprades2018reproducibility}, so the committed engine version is a
real artifact rather than a nominal one. The FIM/CRLB engine itself is Validated
against NumPy and the published bounds of \citet{kay1993}; the lunar application
of it stays Modelled.

Two external-validation steps are scoped and honest about what they would and
would not move. First, tying the engine's lunar-time transform to the open,
IAU-Resolution-compliant LTE440 lunar time ephemeris would let kshana reproduce a
\emph{released} selenocentric-coordinate-time rate against published values---note
that this validates a distinct quantity (a centre-of-mass coordinate-time scale)
and does not turn the surface-clock rate or the joint-OD result into measurements.
Second, the joint-OD observability result will stay Modelled until real lunar VLBI
or ranging data exist; no current dataset can flip it, and we say so rather than
imply otherwise. The exact arXiv identifier, the Zenodo DOI for the v0.23.0 lunar
release, and the repository URL and tag are filled in at posting.

\section*{Acknowledgements / funding}

Ashforde O\"U. No external funding. The author thanks the open-source PNT and
geodesy communities whose public references and conventions this work calibrates
against.

\bibliographystyle{plainnat}
\bibliography{lunar-pnt}

\end{document}